\def\PNout#1{\sout{#1}}

\def\PNout#1{}
\def\diff{\mathrm{d}}

\def\a{\mathrm{a}}

\def\A{\mathrm{A}}
\def\B{\mathrm{B}}
\def\C{\mathrm{C}}
\def\D{\mathrm{D}}

\def\Ek{E_{\textrm{k}}}
\def\vecr{\mathbf{r}}
\def\vecp{\mathbf{p}}
\def\Ntest{N_{\textrm{test}}}

\documentclass[twocolumn,psfig,preprintnumbers,amsmath,amssymb]{revtex4-1}
\usepackage{graphicx}\usepackage{epsfig}
\usepackage{amsmath}\usepackage{amssymb}\usepackage{amsfonts}
\usepackage{bm}
\usepackage{exscale}
\usepackage{dcolumn}    
\usepackage{color}
\usepackage{soul}
\usepackage{ulem}
\begin{document}
%
%
\title{Frustrated fragmentation and re-aggregation in nuclei: \\a non-equilibrium description in spallation}
%
%
\author{P.Napolitani$^1$}
\author{M.Colonna$^2$}
%
%
\affiliation{$^1$ IPN, CNRS/IN2P3, Universit\'e Paris-Sud 11, 91406 Orsay cedex, France}
\affiliation{$^2$ INFN-LNS, Laboratori Nazionali del Sud, 95123 Catania, Italy}
%
%
\begin{abstract}
	Heavy nuclei bombarded with protons and deuterons in the 1 GeV range have a large probability of undergoing a process of evaporation and fission; less frequently, the prompt emission of few intermediate-mass fragments can also be observed.

	We employ a recently developed microscopic approach, based on the Boltzmann-Langevin transport equation,
to investigate the role of mean-field dynamics and phase-space fluctuations in these reactions. 

	We find that the formation of few IMF's can be confused with asymmetric fission when relying on yield observables, 
but it can not be assimilated to the statistical decay of a compound nucleus when analysing the dynamics and kinematic observables: it can be described as a fragmentation process initiated by phase-space fluctuations, and successively frustrated by the mean-field resilience.
	As an extreme situation, which corresponds to non-negligible probability, the number of fragments in the exit channel reduces to two, so that fission-like events are obtained by re-aggregation processes.

	This interpretation, inspired by nuclear-spallation experiments, can be generalised to heavy-ion collisions from Fermi to relativistic energies, for situations when the system is closely approaching the fragmentation threshold.

\end{abstract}



%
%
%
\maketitle
%
%
%
%
%

\section{Context}
	Several decades passed from Serber's early description~\cite{Serber47} of nuclear reactions induced by nucleons and light nuclei at few hundred MeV per nucleon.
	In their standard outline, such reactions, generally called spallation, could be described as a fast excitation of an atomic nucleus, followed by a slower decay process~\cite{Thomas68,Sanders99}: depending on the 
phase space available \cite{Moretto75,Moretto89}, the system undergoes a sequence of more or less asymmetric splits~\cite{Businaro55a,Businaro55b} ranging from particle evaporation to fission.
	It was found already in pioneering studies~\cite{Grover62,Kaufman76,Warwick82,Hirsch84,Andronenko86,Barz86,Korteling90,Kotov95,Hsi97,Avdeyev98}, 
that the most excited systems can also produce intermediate-mass fragments (IMF), and lead to a richer phenomenology comparable with nucleus-nucleus collisions and nuclear fragmentation~\cite{Cole2000,Hufner84,Lynch87,Karnaukhov99}.
	Further research focused on the study of thermodynamic observables from spallation reactions in the relativistic domain~\cite{ISIS,Botvina85b,Botvina90,Karnaukhov03}, in connection with the liquid-gas phase transition in nuclear matter~\cite{Binder1984,Chomaz2000_2003}, and in parallel with the research on the multifragmentation process observed in ion-ion collisions in the Fermi-energy domain~\cite{Bowman1991,Bondorf95,Borderie2008,Moretto2011,Dagostino2000}.

	Several fields of application, from energy and environment to neutron sources and exotic beams, stimulated intense research on protons and deuterons in the 1 GeV range impinging on heavy nuclei.
	The production of some specific light nuclides with large kinetic energy resulted of great relevance in several technical issues (radiation damages, fragilisation of structural materials in accelerator-driven systems, side effects in medical hadron-therapies). This despite the minor contribution of the whole IMF production to the total reaction cross section, which was found to amount to few millibarn.
	In more recent experiments, the possibility of correlating isotopic cross sections to high-resolution kinematic observables allowed tracking the process of the IMF production: it was advanced that it could originate from a melange of two processes, multifragmentation from the most excited configurations~\cite{Napolitani04}, and asymmetric fission~\cite{Ricciardi06}.
	
	This work proposes a fully dynamical description of the process, within a Boltzmann-Langevin transport approach, with the aim of probing the mechanism which rules the physical process,
through the analysis of kinematic and correlation observables.


\section{The quest on the origin of IMF\\ in spallation}
	The IMF production in spallation, especially in the 1$A$ GeV range, is a process at the threshold between multifragmentation and compound-nucleus decay.
	To characterise the process, an ideal experimental approach should measure event-by-event particle-particle and kinematic correlation observables at high-resolution.
	This goal has been only partially achieved, so that the lack of resolution or the missing of some correlations opens the way for a variety of physical interpretations, ranging from attributing all IMF's to sequential 
fission processes to the opposite extreme that all IMF's signal multifragmentation events.
	In the following we will refer to two specific experimental approaches, implying that the same considerations could follow from other experimental analyses.

\subsection{Experimental observables: an example}

\begin{figure}[b!]\begin{center}
	\includegraphics[angle=0, width=1\columnwidth]{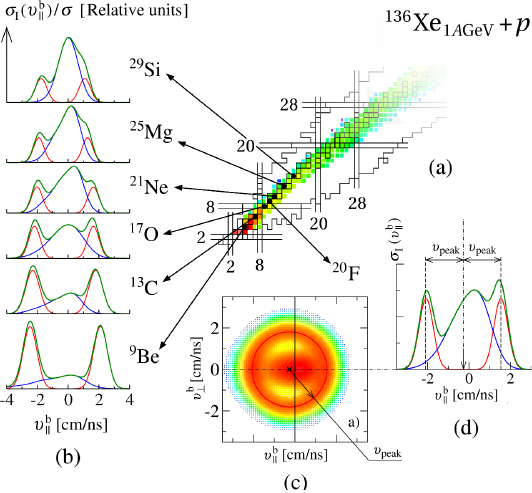}
\end{center}
\caption{
	(a) Isotopic cross sections $\sigma$ measured in the IMF region for the system $^{136}$Xe$+p$ at 1$A$GeV, from the experiment of ref.~\cite{Napolitani07}.
	(b) Zero-angle invariant velocity distributions $\sigma_{\textrm{I}}(v_{||}^{\textrm{b}})/\sigma$ extracted from the experimental data for some isotopes~\cite{Napolitani2011}. 
	(c) Analysis detail for one isotope, $^{20}$F: invariant velocity plot
on a plane containing the longitudinal axis. Two contributions are present with a recoil mismatch: a Coulomb ring with radius $v_{\textrm{peak}}$ and a gaussian-like distribution.
	(d) Reduction of the two-dimensional plot to a distribution along the
beam axis: two kinematic contributions to the invariant-velocity spectrum appear as a convex mode (Coulomb ring) and a concave mode (gaussian-like), respectively.
}
\label{fig_v_distr}
\end{figure}

	The first experimental approach we focus on is a campaign of inverse-kinematics experiments, performed at GSI (Darmstadt): the FRagment Separator~\cite{Geissel92} (FRS), a high-resolution magnetic achromat~\cite{Schmidt87}, was employed to measure the nuclide production in several spallation reactions at around 1$A$GeV~\cite{Wlazlo2000,Benlliure01,Rejmund01,Enqvist01,Enqvist02,Taieb03,Bernas03,Armbruster04,Bernas06,Casarejos06,Pereira07,Napolitani04,Villagrasa07,Napolitani07,Benlliure08}.
	In some systems it was possible to measure the isotopic cross sections of the IMF and the corresponding zero-angle invariant-velocity distributions; these distributions are constructed by selecting only the velocity vectors aligned along the beam direction (they are evidently different from longitudinal projections of the whole velocity distributions).
	Fig.~\ref{fig_v_distr} presents some of those experimental results taken from ref.~\cite{Napolitani07}, and resumes few essential steps of the data analysis~\cite{Napolitani2011}.
%
%
%

	As evident in Fig.~\ref{fig_v_distr}, in the IMF region the velocity spectra of a given nuclide is the composition of two shapes: one is concave (showing two peaks), the other is convex.
	The two contributions were also found to have different relative shares as a function of the nuclide.	
	Though, the presence of two different contributions is evident only in the kinematics, while the nuclide production selected for either the concave or the convex mode contribute to the same region of the nuclide chart, in general situated in the neutron-rich side with respect to $\beta$ stability: this is the reason why the presence of these two modes was invisible in many experiments.
Moreover, the two contributions are associated with a shift in the mean value of the spectra, indicating incompatible values for the mean momentum transfer: this latter reveals in fact the violence of the entrance channel.

	The concave shape is reflected in a Coulomb-shell velocity distribution probed at zero angle.
	The radius of the shell is compatible with a fission kinematics ~\cite{Beck96}
and the mean value evolves coherently with empirical systematics for the mean recoil momentum as a function of the fissioning system (i.e. compatible with the systematics of Morrissay~\cite{Morrissey89}).

	The convex shape is one broad hump, often asymmetric, which signs 
the folding of many different contributions; it is associated with a mean recoil momentum which does not follow any empirical systematics.
	This indicates that the fissioning configuration is not achieved, either because the multiplicity of fragments is not equal to two, or because the kinematics is not consistent with a conventional fission configuration.
	In this respect, the convex shape is rather compatible with multifragmentation.

	Such observations led to the conclusion that the IMF production should combine asymmetric-fission (concave shapes) and multifragmentation (convex shapes).
	On the one hand, this description was rigorously established because these two velocity contributions were measured at the same time for each single nuclide~\cite{Napolitani2011}.
	On the other hand, the problem of assuring such interpretation was that particle-particle correlations and IMF multiplicities were not measured in the inclusive approach.
	Especially for the convex shape this information is important as multifragmentation is usually associated with a large number of IMF of similar size.

	A second experimental technique was adapted to obtain this information.
	Some of the systems previously measured inclusively at the FRagment Separator were successively measured again with an exclusive approach in the Spalladin experimental campaign~\cite{LeGentil08,Gorbinet2011,Gorbinet2012} at GSI (Darmstadt).
	These experiments indicated that IMF are observed in events where fragment multiplicity is prevalently equal to two, and that events with larger multiplicity were more rare.
	On the one hand, this confirmed that there are actually two contributions to the IMF production, a binary channel and a higher multiplicity channel. 
	On the other hand, the events exceeding two IMF were not easily identifiable with ordinary multifragmentation because of the low IMF multiplicity and the size asymmetry.
	This encouraged interpretations fully relying on statistical models, where the IMF production is obtained either from a sequence of (asymmetric) fission contributions, or sampled from an ensemble of possible multifragmented configurations.
	In general, these approaches are both an efficient workaround because they define directly the outcome of the reaction on the basis of the involved excitation energy, and they can yield quite similar results for the IMF production despite implying different physical pictures.
%
	Then the question arises whether we can include spallation in the multifragmentation picture.

	It is our intention to illustrate that the key to achieve a fully coherent understanding of this apparently self-contradictory experimental information, conciliating the possibility to emit only few IMF's
with the observation of new kinematic properties,
 is a microscopic dynamical description of the process.

\subsection{Need of a dynamical description\\ to address the quest}
	Avoiding prominent mechanical contributions from the entrance channel which characterise ion-ion collisions at Fermi energies, spallation is a favoured processes to produce a thermalised remnant.
	It is therefore well established that the usual hypothesis of a hot and fully equilibrated source is perfectly adapted to spallation and in fact statistical models proceeding from this assumption are well suited to reproduce large part of the experimental observables~\cite{Napolitani04}.
	On the other hand, it is not the purpose of these approaches to investigate the reaction mechanism which leads to the final configuration. 

	To directly address the quest, we proceed through a dynamical description avoiding any a priori assumption on the degree of equilibration of the system at a given reaction time.

	In particular, the dynamical approach allows to describe the possible onset of mechanical instabilities in a fermionic system: this is a general process which characterises Fermi liquids at low densities and which results in inhomogeneous density patterns.
	As we will explore in the following, the incident energy of a light projectile can actually be large enough to produce the thermodynamical and density conditions for unstable modes to get amplified, i.e. to enter the nuclear spinodal region~\cite{Chomaz2004}.
	The dispersion relation~\cite{Colonna1994} defines how the growth rate of these unstable modes connects to the mean-field potential.
	In this direction, a one-body approach based on an efficient description of the dispersion relation is well adapted, and a stochastic treatment is necessary to sample the variety of possible dynamical trajectories that unstable conditions may produce.
	With the purpose of investigating fragment formation, stochastic one-body approaches have already been applied to spallation~\cite{Colonna1997}, but with incomplete success because fluctuations were not treated in full phase-space.
	As an improvement, in this work we employ the Boltzmann-Langevin One Body (BLOB) approach~\cite{Napolitani2013} which has been constructed under the explicit constraint of describing the fluctuation mechanism in full phase-space by solving the Boltzmann-Langevin (BL) transport equation in three dimensions.
	The model was firstly applied to dissipative central ion-ion collisions~\cite{Napolitani2013,Napolitani2014} at Fermi energies, to investigate the transition from incomplete fusion to multifragmentation in proximity of the low-energy threshold for multifragmentation, which we expect spallation can also probe in the 1$A$GeV regime.

\section{Boltzmann-Langevin One Body description of a spallation system}
	Within the BLOB transport model we follow the dynamics of the target nucleus, after the interaction with the light projectile.
	We describe the $N$-body system with a one-body Hamiltonian $H$ supplemented by a fluctuating contribution to account for the unknown $N$-body correlations.
	The BL equation describes the time evolution of the semiclassical one-body distribution function  $f(\vecr,\vecp,t)$ in its own self-consistent mean field:
\begin{equation}
	\partial_t\,f - \left\{H[f],f\right\} = {\bar{I}[f]}+{\delta I[f]} \;.
\label{eq1}
\end{equation}
The left-hand side gives the Vlasov evolution for $f$ 
and the right-hand side introduces the residual interaction, which also carries the unknown N-body correlations.
	This latter contains the average Boltzmann hard two-body collision integral 
$\bar{I}[f]$ and the fluctuating term of Markovian type $\delta I[f]$, also written in terms of the one-body distribution function \cite{Ayik1990}.

	The propagation of the one-body distribution function is described through the test particle method and employs a Skyrme-like (\textit{SKM}$^*$) effective interaction~\cite{Guarnera1996}, defined according to a soft isoscalar equation of state (of compressibility $K\!=\!200$~MeV); the potential energy per nucleon $E_{\textrm{pot}}/A$ is defined as
\begin{equation}
	\frac{E_{\textrm{pot}}}{A}(\rho) = 
		\frac{A}{2}u
		+\frac{B}{\sigma+1}u^\sigma
		+\frac{C_{\textrm{surf}}}{2\rho}(\nabla\rho)^2
		+\frac{1}{2}C_{\textrm{sym}}(\rho)u\beta^2 ,
\label{eq_pot}
\end{equation}
with $u=\rho/\rho_0$ and $\beta=(\rho_n-\rho_p)/\rho$,
where $\rho_0$ and $(\rho_n-\rho_p)$ are the saturation and the isovector density, respectively.
The parameters are $A\!=\!-356$ MeV, $B\!=\!303$ MeV, $\sigma\!\!=\!7/6$. 
	Surface effects are accounted for by considering finite width wave packets
for the test particles employed in the numerical resolution of Eq.(1).
The explicit term added to the potential energy is tuned to reproduce the surface energy of nuclei in the ground state \cite{Guarnera1996}.
A linear (stiff) density dependence of $E_{\textrm{pot}}$ is considered by choosing $C_{\textrm{sym}}(\rho)\!=\!{\textrm{constant}}\!=\! 32$MeV.

	The fluctuating term $\delta I[f]$ acts as a dissipating force 
during the whole temporal evolution of the process and introduces fluctuations by exploiting $N$-body correlations.
In the BLOB procedure, 
binary collisions 
involve extended phase-space agglomerates of test particles of equal isospin A$={a_1,a_2,\dots}$, B$={b_1,b_2,\dots}$ to simulate nucleon wave packets:
\begin{equation}
	{\bar{I}[f]}+{\delta I[f]}
	= g\int\frac{\diff\vecp_b}{h^3}\,
	\int \diff\Omega\;\;
	W({\scriptstyle\A\B\leftrightarrow\C\D})\;
	F({\scriptstyle\A\B\rightarrow\C\D})\;,
\label{eq2}
\end{equation}
where $g$ is the degeneracy factor, $W$ is the transition rate, in terms of relative velocity between the two colliding agglomerates and differential nucleon-nucleon cross section
\begin{equation}
	W({\scriptstyle\A\B\leftrightarrow\C\D}) = |v_\A\!-\!v_\B| \frac{\diff\sigma}{\diff\Omega}\;,
\label{eq3}
\end{equation}
and $F$ contains the products of occupancies and vacancies of initial and final states calculated for the test-particle agglomerates
\begin{equation}
	F({\scriptstyle\A\B\rightarrow\C\D}) =
	\Big[(1\!\!-\!\!{f}_\A)(1\!\!-\!\!{f}_\B) f_\C f_\D - f_\A f_\B (1\!\!-\!\!{f}_\C)(1\!\!-\!\!{f}_\D)\Big]\;.
\label{eq4}
\end{equation}
	
	Since $\Ntest$ test particles are involved in one collision, and since those test particles could be sorted again in new agglomerates to attempt new collisions in the same interval of time as far as the collision is not successful, the nucleon-nucleon cross section contained in the transition rate $W$ should be divided by $\Ntest$: $\sigma = \sigma_{\textrm{NN}} / \Ntest$.
	In this work the $\sigma_{\textrm{NN}}$ is taken equal to the free nucleon-nucleon cross section, with a cutoff at $100$~mb.	Moreover, the differential cross section depends on the scattering angle according to the prescription of ref.~\cite{Bertsch1988}.

	All test-particle agglomerates are redefined at successive intervals of time in phase-space cells of volume $h^3$; in their initial state they correspond to the most compact configuration in the phase-space metrics which does neither violate Pauli blocking in the initial and in the final states, nor energy conservation in the scattering.
	The metrics of the test particle agglomerates is defined in such a way that the packet width in coordinate space is the closest to $\sqrt(\sigma_{\textrm{NN}}^{\textrm{medium}}/\pi)$, where  $\sigma_{\textrm{NN}}^{\textrm{medium}}$ corresponds to the screened cross section prescription proposed by Danielewicz~\cite{DanielewiczCoupland}, which was found to describe recent experimental data ~\cite{Lopez2014}. In this way, the spatial extension of the packet decreases as the nucleon density increases. 
	The nucleon-nucleon correlations produced through this approach are then exploited within a stochastic procedure, which consists in confronting the effective collision probability $W\times F$ with a random number.
	When the scattering is successful, a precise shape-modulation technique~\cite{Napolitani2012} is applied to ensure that the occupancy distribution does not exceed unity in any phase-space location in the final states.
	Such constraint avoids that the Pauli blocking could be violated, and it imposes to pay special attention to the metrics of the phase space (see discussion in ref.~\cite{Chapelle1992}).

	As a consequence, fluctuations develop spontaneously in large portions of phase space, with an amplitude variance equal to $f(1\!-\!f)$, at equilibrium,  in a phase-space cell of volume $h^3$.	
This leads to a correct Fermi statistics for the distribution function $f$, in terms of mean value and variance.
	Calculations in a periodic box for unstable nuclear matter, in one dimension~\cite{Rizzo2008} 
and in three dimensions~\cite{Napolitani_IWM2014} have shown that the BLOB approach describes the growth rate of the corresponding (spinodal) unstable modes, related to the form of the mean-field potential, as ruled by the dispersion relation. 
	Thus the BLOB model for heavy-ion collisions is constructed as based on the efficient description of these aspects. 

\subsection{Definition of the heated system}
	Differently from heavy-ion collisions at Fermi energy, in this application to spallation only the dynamics of the heated heavy nucleus is followed after a suitable initialisation.
	As usual, the system is initially defined by organising the test particles in a minimum-energy configuration in accordance with the form chosen for the nuclear interaction.
	In order to define the heated system, this configuration is redefined by processing a simplified cascade induced by the incoming light projectile: the amount of energy deposited by the projectile in traversing the nucleus is calculated as well as the corresponding distribution in phase space.
	A time-dependent calculation would require very small time steps and a relativistic formalisation of the dynamics.
	Due to the rapidity of the spallation process with respect to the dynamics of the heated system, it is convenient to reduce the cascade to an approximate description where only test-particles from the incoming relativistic projectile are followed along space trajectories and target test-particles are not displaced during the cascade.
	In practice, this simplification is made by reducing the cascade process to a calculation of the energy loss of the projectile, modifying the momentum landscape of the target test-particles without processing any time evolution of the system in coordinate space.
	For relativistic projectiles this choice is not incompatible with the observation that the projectile leaves the target nucleus before that the swarm of the first fastest ejectiles appears at the surface of the target nucleus~\cite{Cugnon1981,Cugnon87}.

%
%
\begin{figure}[b!]\begin{center}
	\includegraphics[angle=0, width=.7\columnwidth]{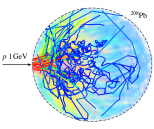}
\end{center}
\caption{
	Energy-deposition map with one bunch of spacial cascade trajectories in relief, calculated for a central impact parameters in the reaction $p+^{208}$Pb at 1$A$GeV.
	Each trajectory is associated to one test particle of the incoming projectile.
}
\label{fig_cascade}
\end{figure}
	Fig.~\ref{fig_cascade} shows a bunch of spacial cascade paths corresponding to a $^{208}$Pb target nucleus bombarded by 1 GeV proton projectiles with a central impact parameter;
the resulting excitation energy distribution corresponds therefore to the most violent events.
	The test particles composing the projectile hit the nucleus along the projectile direction within a cylinder of radius equal to the projectile radius.
	Each projectile test particle produces a cascade path inside of the target nucleus, which is redefined at each scattering occurrence:  after scattering, the projectile test-particle trajectory continues to be followed along the fastest scattered particle and the other particle, after being assigned a new momentum, is no more followed.
	Between two scattering points the path is a straight segment.
	All cascade paths traced by the projectile test particles are followed in coordinate space simultaneously. 
	For a couple of target and projectile test particles, the collision is searched according to the closest-approach criterion applied to the corresponding centre-of-mass energy $\sqrt{s}$~\cite{Bertsch1988} and by using the same nucleon-nucleon cross section used for the transport calculation.
	All collisions are considered as elastic scatterings;  
the model could be improved by including the $\Delta$ production-absorption mechanism, but we consider the present simplified treatment sufficient for the purpose of obtaining the excitation energy of the target nucleus.
	A strict Pauli-blocking condition here is imposed, so that only scattering events which create a hole and a particle outside of the Fermi sphere are accepted; otherwise, the target test particle could participate to a scattering with another projectile test-particle.
	When the cascade trajectories hit the inner potential boundary of the system, they can traverse the boundary according to the corresponding transmission probability, calculated with the relativistic formalisation proposed in ref.~\cite{Cugnon1997}; the potential depth used for calculating the transmission probability is 40MeV, which represents the average value characterising the bulk of the system.
	This transmission probability is used to calculate an additional portion of the total energy of the projectile, which is considered dissipated in the target system and which corresponds to the reflected wave.

%
%
\begin{figure}[b!]\begin{center}
	\includegraphics[angle=0, width=1\columnwidth]{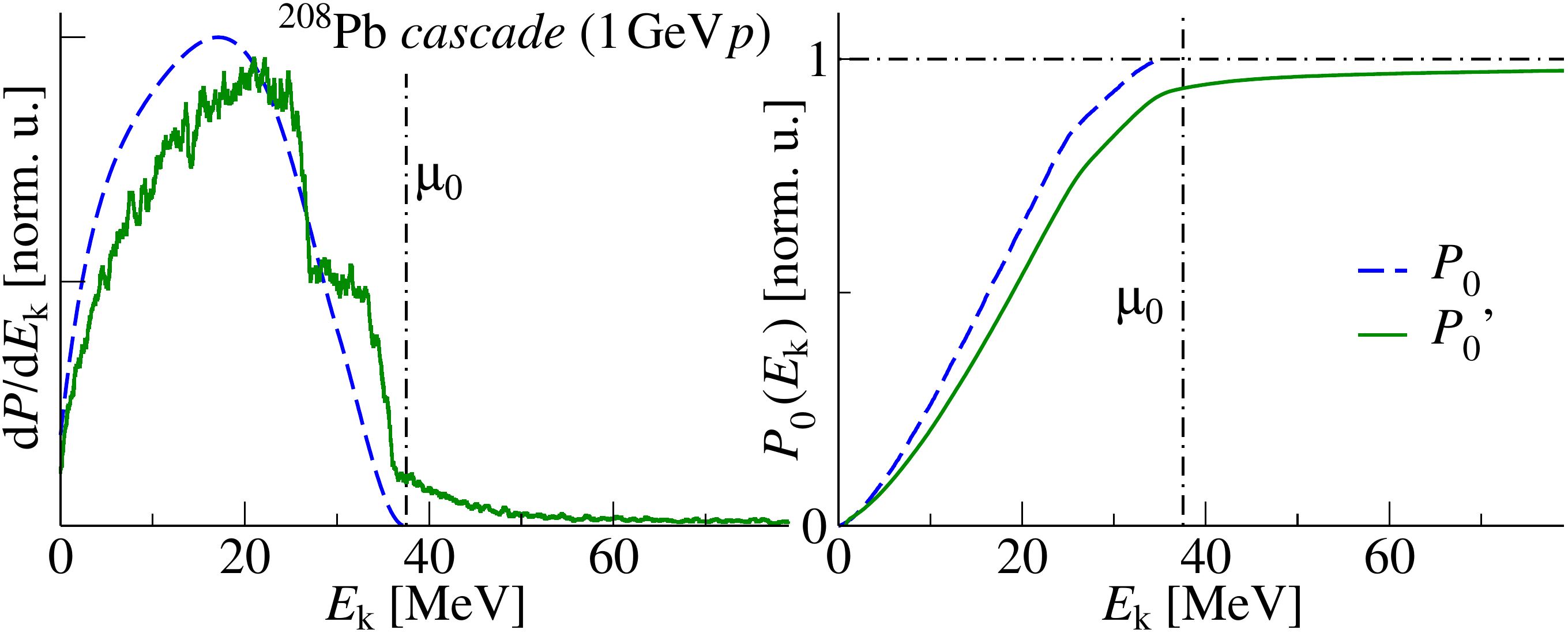}\\
\end{center}
\caption{
	Effect of the cascade in modifying the momentum space from the initial state configuration (dashed line)
to the excited configuration (full line), calculated for one event in the reaction $p+^{208}$Pb at 1 $A$GeV 
(see text).
	The panel on the left represents the nucleon energy distribution, whereas on the right the integrated energy distribution is presented.
}
\label{fig_finalstate}
\end{figure}
%
%
%
%
\begin{figure}[t!]\begin{center}
	\includegraphics[angle=0, width=.95\columnwidth]{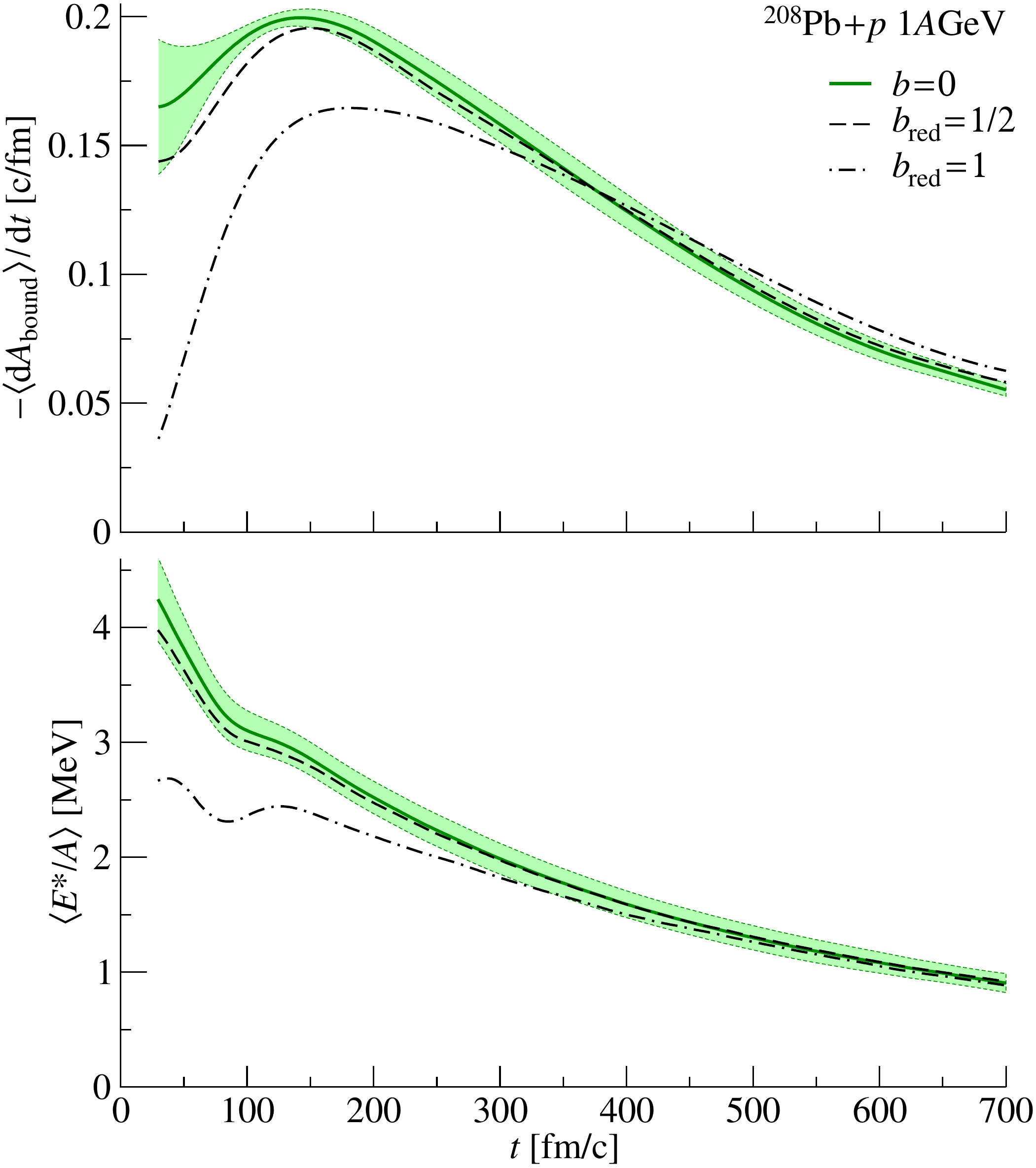}
\end{center}
\caption{
	Upper panel. Evolution of the mean number of emitted nucleons per interval of time in the reactions $p+^{208}$Pb at 1 GeV.
	Lower panel. Evolution of the mean excitation energy calculated for the fraction of bound matter in the reaction $p+^{208}$Pb.  
	Central, intermediate and peripheral impact parameters are tested.
The width of the bands give the standard deviation around the trajectories for the central collisions; other trajectories have a comparable standard deviation (not indicated).
}
\label{fig_collcasc}
\end{figure}
	While the coordinate space is frozen to its initial configuration, the initial momentum distribution is updated according to the cascade scatterings.
	The energy deposited by the projectile in the system is then obtained by considering the momentum variation, supplemented by the reflection contribution at the potential boundary of the system.	
	Accordingly, as shown in fig.~\ref{fig_finalstate} for the reaction $p+^{208}$Pb at 1 GeV, the initial integrated kinetic energy distribution $P_0(\Ek)$ is modified into a new distribution $P_0'(\Ek)$.

	The cooling process of the excited system is then followed in time with BLOB. 
The latter is shown in fig.~\ref{fig_collcasc}, setting the reduced impact parameter $b_{\textrm{red}}$ (impact parameter divided by the target radius) equal to 0, 0.5 and 1 for the system $p+^{208}$Pb at 1 GeV.
	The evolution of the mean fraction of bound matter $\langle \diff A_{\textrm{bound}}\rangle / \diff t$ tracks the mean number of emitted nucleons per interval of time: central and intermediate impact parameters act almost equally in removing a large part of nucleons, while peripheral collisions favour the formation of heavier remnants.
	The corresponding information is carried by the evolution of the mean excitation energy per nucleon $\langle E^*/A\rangle$ averaged over all portions of bound matter in the system.

\section{Nuclide production \\and kinematics}
	The model described above was applied to six systems, chosen because close to some significant experimental data and because they constitute a series of successive variations of only one parameter among projectile, target and energy:
$^{208}$Pb+$p$ at 1 $A$ GeV,
$^{208}$Pb+$p$ at 750 $A$ MeV,
$^{208}$Pb+$d$ at 750 $A$ MeV,
$^{197}$Au+$d$ at 750 $A$ MeV,
$^{136}$Xe+$p$ at 1 $A$ GeV and
$^{124}$Xe+$p$ at 1 $A$ GeV.
%
%
\begin{figure}[b!]\begin{center}
	\includegraphics[angle=0, width=1\columnwidth]{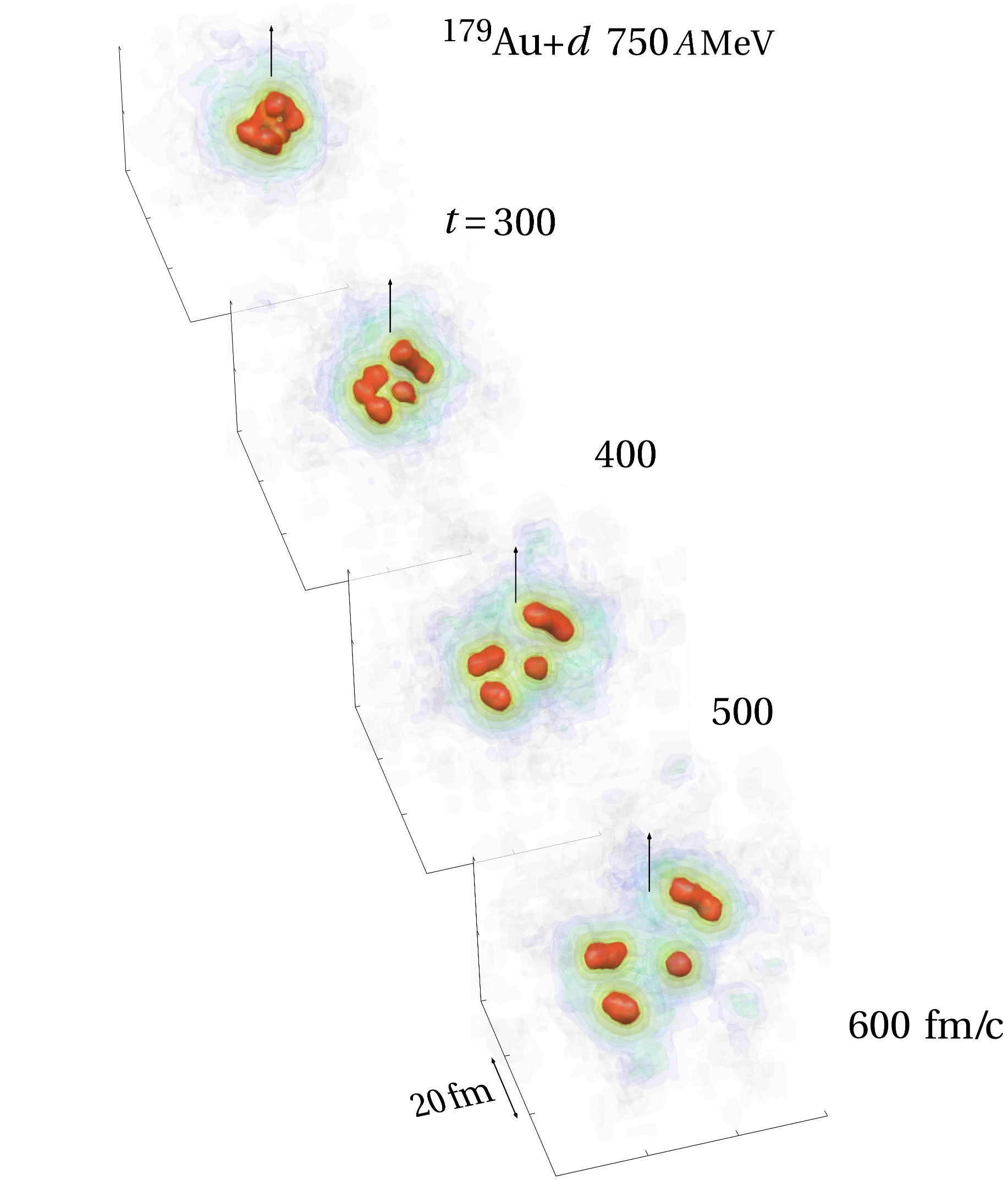}
\end{center}
\caption{
	Study of the fragment configuration for one event of the system $^{197}$Au+$p$ at 750 $A$ MeV, for reaction times which are compatible with the definition of the fragmentation pattern.
The event shown in the figure is selected among those giving the largest fragment multiplicity.
}
\label{fig_animation_exit}
\end{figure}
	The dynamical calculations were performed reducing to central impact parameters in the range $0<\!b\!<0.75$fm, with the purpose of restricting to the small portion of geometric cross section where the contribution of heavy residues is not dominant, and where IMF formation is enhanced. 
	 The remaining fraction of cross section favouring compound-nucleus decays can be efficiently described through statistical approaches.
	Such a choice is however schematic because, due to fluctuations in the cascade trajectories, the impact parameter is not directly characterising the entrance channel, and violent collisions may arise also in less central configurations with smaller probability.
	Conversely, less excited configurations are also associated to central impact parameters with smaller proportion than in peripheral collisions.
	A statistics of about 1500 
stochastic events per system have been collected, using a 32 CPU parallel computing station.

\subsection{Dynamical description up to the formation of primary fragments}

%
%
\begin{figure}[b!]\begin{center}
	\includegraphics[angle=0, width=.9\columnwidth]{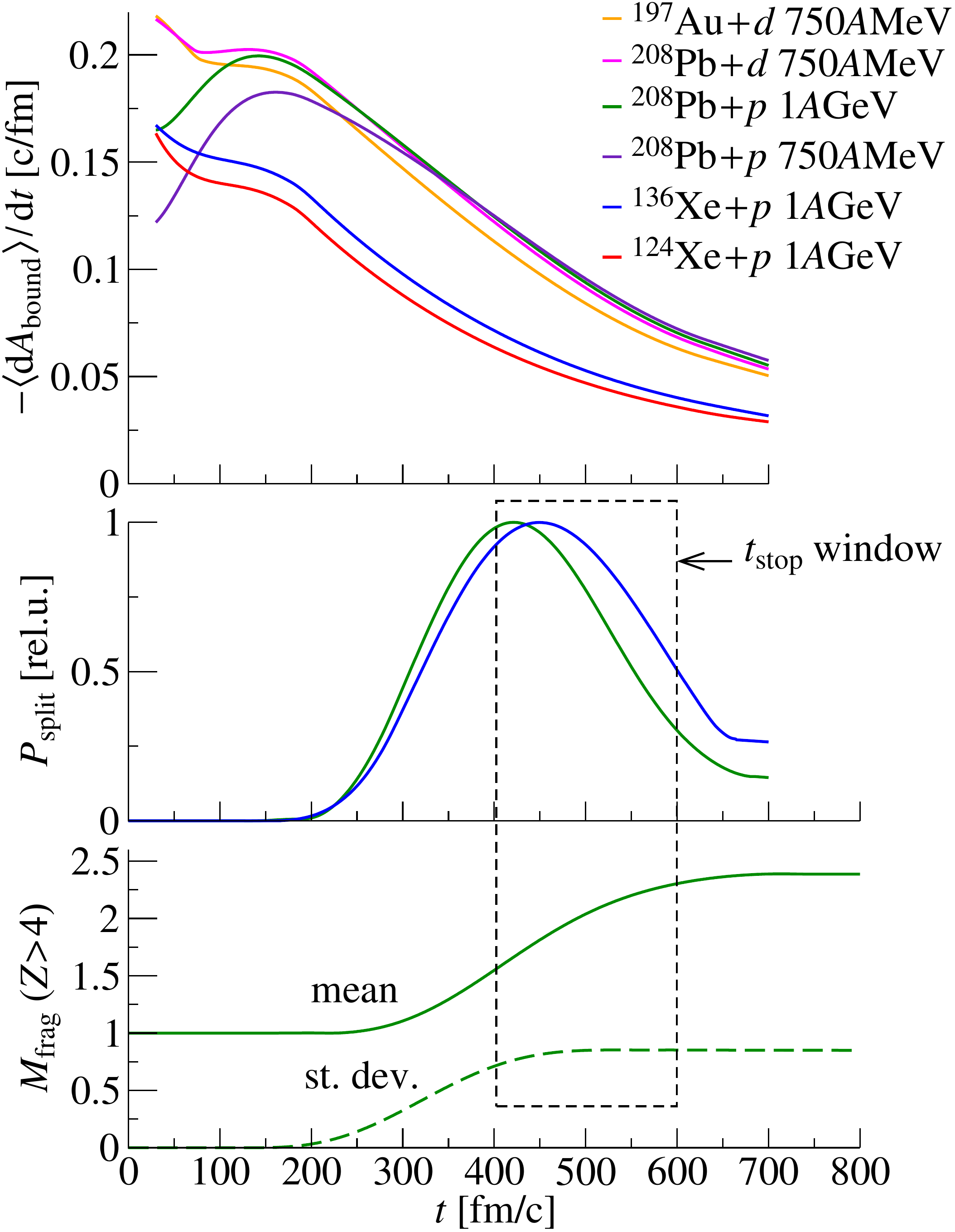}
\end{center}
\caption{ 
	Upper panel. As in fig.~\ref{fig_collcasc}: evolution of the mean number of emitted nucleons per interval of time for the spallation reactions described in the text.
Middle panel. Probability of split as a function of time for  $p+^{208}$Pb and $p+^{136}$Xe at 1GeV.
Lower panel. Saturation of the fragment multiplicity $M_{\textrm{frag}}$ for $p+^{208}$Pb at 1GeV; Mean and standard deviation are shown for the multiplicity of fragment with $Z\!>\!4$.}

\label{fig_mass_loss}
\end{figure}

	Within the model described above, in fig.~\ref{fig_animation_exit} we represent one possible evolution of the density profile of the systems $^{197}$Au+$p$ at 750 $A$ MeV for a central impact parameter; this is a rather rare event corresponding to the fragmentation of the target nucleus in more than three fragments.
	The system breaks into three asymmetric parts visible at 400~fm/c.
At later times, further splits may proceed from some individual largely deformed sources, as displayed in fig.~\ref{fig_animation_exit} for the time 500~fm/c.
	In these spallation processes the fragment multiplicity saturates after 700~fm/c.
	This is shown in fig.~\ref{fig_mass_loss} (bottom), in correlation with the particle emission and the corresponding reduction of bound mass as a function of time for all the simulated systems (top).
The middle panel of the figure shows the probability of observing a split in the system, as a function of time, for two
of the reactions considered. 
 
\begin{figure}[t!]\begin{center}
	\includegraphics[angle=0, width=1\columnwidth]{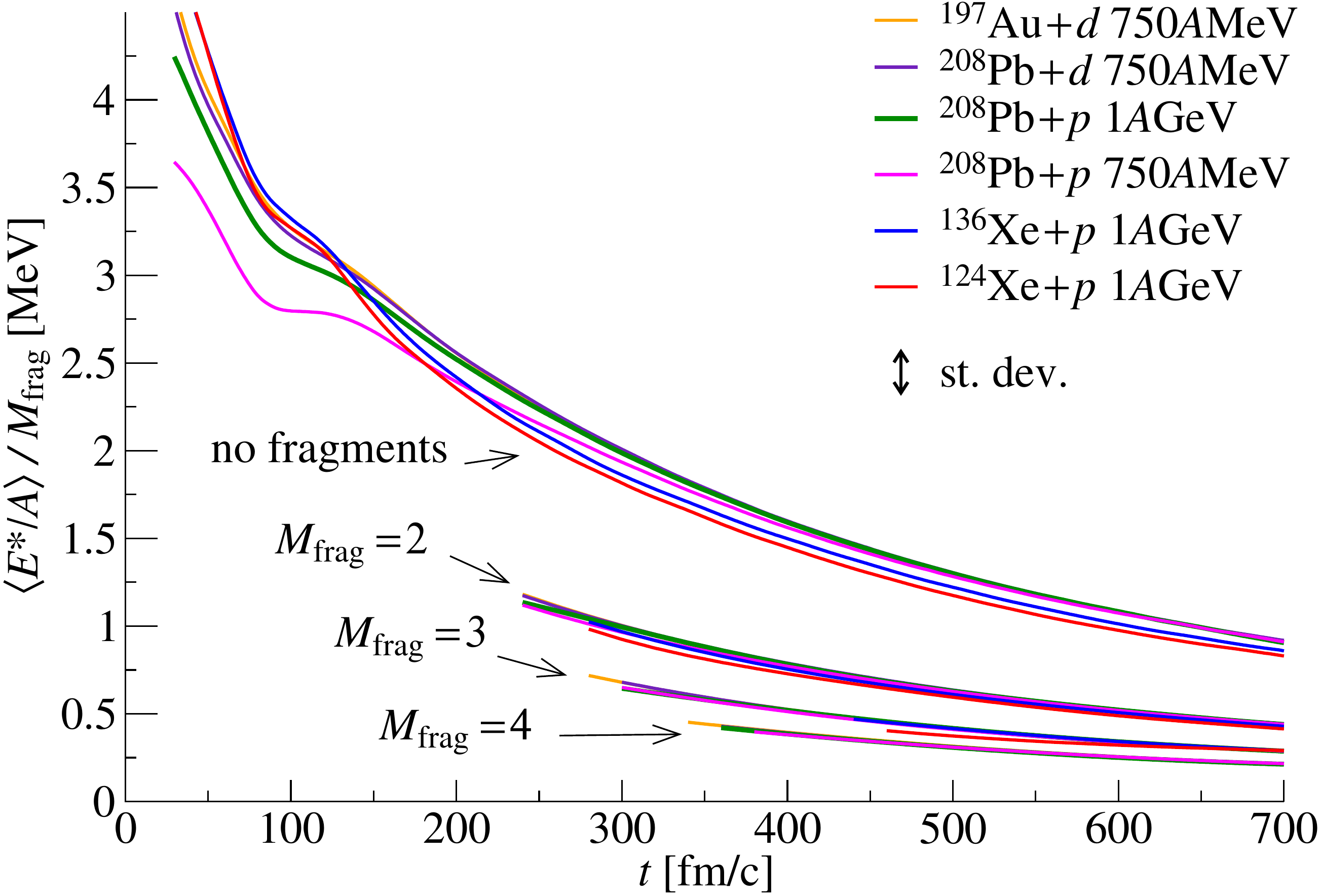}
\end{center}
\caption{
	Evolution of the average excitation energy $\langle E^*/A\rangle$ for the fraction of bound matter during the reaction. The double arrow gives the average uncertainty in terms of standard deviation.
	The bunch of lines extending over the whole time evolution describes events where only a heavy residue is present; bunches of lines for systems which split into $M_{\textrm{frag}}$ equal to two, three or four fragments are divided by $M_{\textrm{frag}}$ for better visibility as they would all collapse on the line for $M_{\textrm{frag}}$=1.
}
\label{fig_energy_time}
\end{figure}

	The cooling process is reflected in the decrease of the average thermal excitation energy per nucleon shown in fig.~\ref{fig_energy_time}.
	A backbending appearing between around 50 and 100 fm/c indicates the attempt of the system to revert the initial pure expansion dynamics into the mechanism of fragment formation.  Indeed, in presence of instabilities, it is energetically convenient for the system to break up into fragments.   
	This also causes a slight increase of the temperature and thus of the thermal excitation energy.
	Event by event, we consider as freeze-out time, $t_{\textrm{stop}}$, the instant between $t\!=\!400$~fm/c and $t\!=\!600$~fm/c where the last split has occurred. Our choice is motivated by the fact that at $t\!\approx\!400$~fm/c the split 
probability is maximum, whereas at $t\!\approx\!600$~fm/c it reduces to a quite low constant value.
For events where a residue is observed, we adopt $t_{\textrm{stop}}\!=\!400$~fm/c. 
	Beyond the time $t_{\textrm{stop}}$ the decay process slows down and only sequential binary splits become possible, which can be efficiently described through a transition-state model. 
	The dynamical calculation is therefore completed with the model SIMON~\cite{Durand1992}, 
which incorporates in-flight Coulomb repulsion.
	
%
%

\subsection{On the way to the residue corridor}

	In correlation with the excitation energy, also the isospin content of fragments and residues evolves in time.
	In general, when a compound nucleus is formed, its excitation energy is extinguished in an attempt of balancing proton and neutron decay widths, so that the bound matter of the systems tends to accumulate along the residue corridor~\cite{Charity98}, which is located in the neutron-deficient side of the nuclide chart with respect to beta stability, and any further decay occurs only along this line in average.
	If however part of the excitation energy is spent in fragmenting the system, neutron rich fragments stop their decay path before reaching the residue corridor, in locations of the nuclide chart which are closer to beta stability, or that are even neutron rich~\cite{Napolitani2011}. 

\begin{figure}[t!]\begin{center}
	\includegraphics[angle=0, width=1\columnwidth]{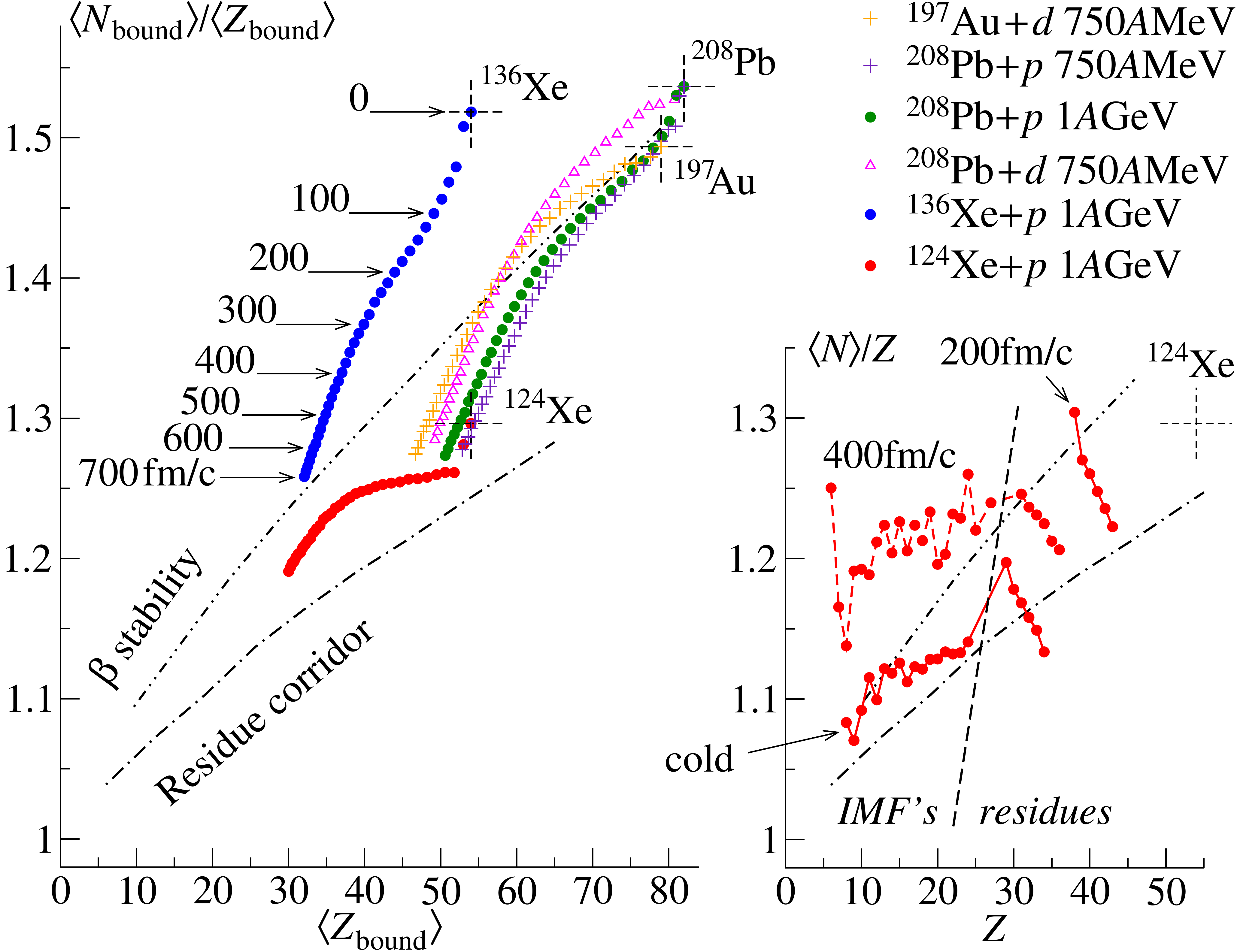}\\
	\includegraphics[angle=0, width=1\columnwidth]{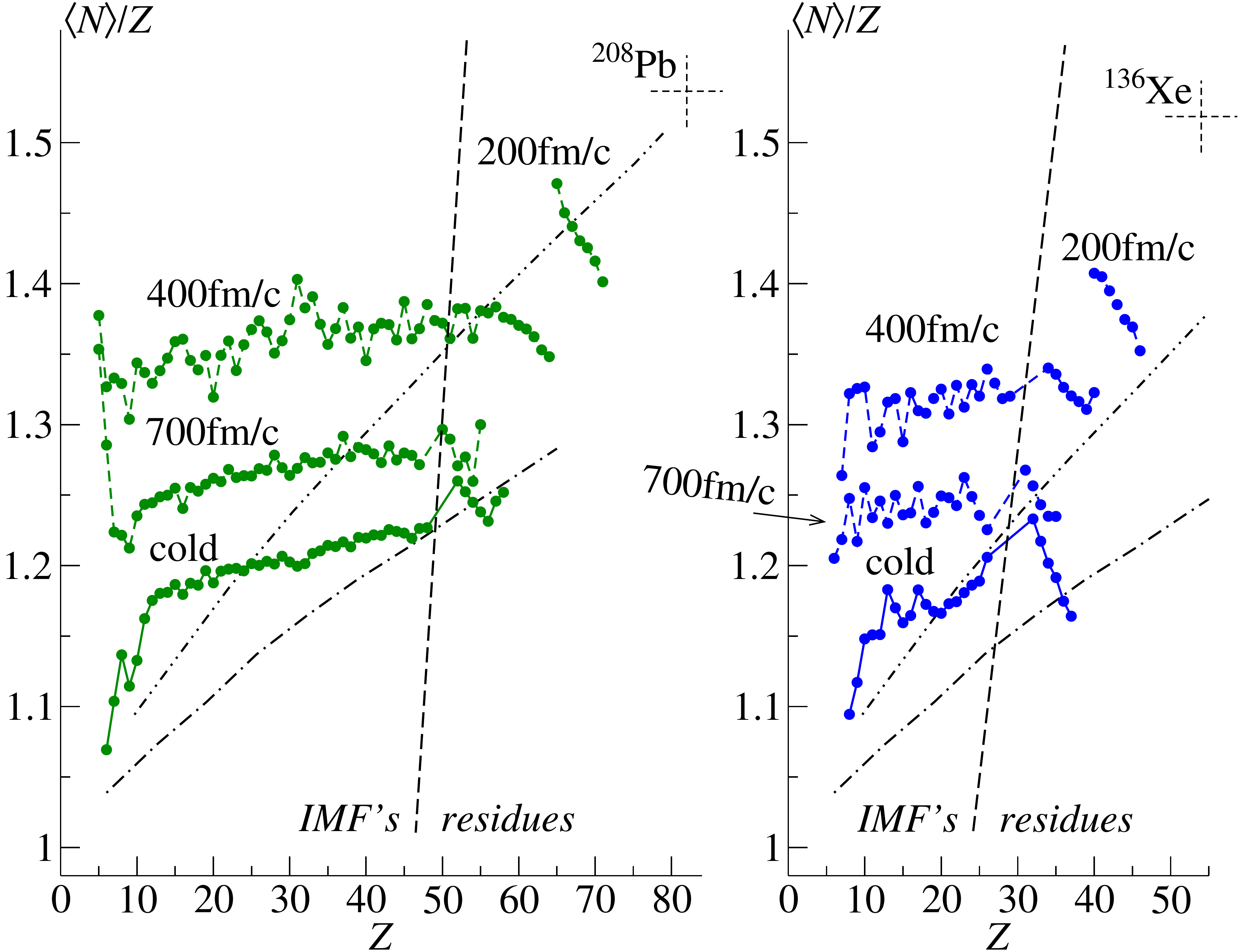}\\
\end{center}
\caption{
	Top left. Evolution of the average isotopic content of bound matter constituting different hot systems as a function of time (calculated for central impact parameters).
	Top right, bottom right, bottom left:
	Distributions of the average isotopic content of single elements produced in the systems 
$^{124}$Xe, $^{136}$Xe and $^{208}$Pb (moving clockwise) bombarded by 1 GeV protons as a function of the element number at 200fm/c (before fragmentation), at 400fm/c (latgest fragmentation probability), at 700fm/c (when the fragment multiplicity saturates in the dynamical calculations) and after secondary-decay progressing from $t_{\textit{stop}}$.
	The $\beta$-stability and the residue corridor are indicated (see text). 
Residues and IMF regions are also indicated.
}
\label{fig_NoverZ}
\end{figure}

	This process inspired several experiments and simulations with statistical models where an assumption of thermal equilibrium of the system was imposed and a temperature was assigned~\cite{Schmidt2002} (the so-called `limiting temperature for fragmentation', corresponding to about 5~MeV).
	The dynamical approach handles this process without any hypothesis of equilibrium.
	Fig.~\ref{fig_NoverZ} (top left panel) examines the evolution of the isotopic content for the six different hot systems for central impact parameters: the average isotopic content of bound matter, obtained by dividing the average number of bound neutrons constituting the system $N_{\textrm{bound}}$ by the average bound charge $Z_{\textrm{bound}}$ is tracked as a function of time until 700~fm/c. 	
	In this interval of time the path moves in average in the direction of the residue corridor while removing mass.
	The whole distribution of the isospin content $\langle N\rangle/Z$ of hot fragments is given in fig.~\ref{fig_NoverZ} for $^{208}$Pb+$p$, for the neutron-deficient system $^{124}$Xe+$p$ and for the neutron-rich systems $^{136}$Xe+$p$ 
(top right, bottom right and bottom left panels); the following times are analysed: $t\!=\!200$~fm/c, before fragmentation, $t\!=\!400$~fm/c, after fragmentation, and $t\!=\!700$~fm/c, when the fragment multiplicity saturates.
	In all the three systems, the distribution at 400~fm/c covers the region of neutron-rich nuclei as a flat function of the element number and its distance from the residue corridor depends on the isospin content of the target nucleus; it drops to smaller values of $\langle N\rangle/Z$ for later times.
	As a function of the available excitation (i.e. of the time), the corresponding distribution of cold fragments is found displaced in the direction of the residue corridor.
	The cold IMF's align along the residue corridor only for the neutron poor target $^{124}$Xe; they do not reach it completely for the neutron rich target $^{136}$Xe, ending their decay path in the vicinity of $\beta$-stability; in the case of the heavy neutron rich target $^{208}$Pb, only the largest fragments can reach the residue corridor at the end of their (shorter) decay path. 
	A complete study of this reaction, involving also less violent events for peripheral impact parameters would extend the distribution of residues to larger mass numbers which would accumulate along the residue corridor.
	The same behaviour characterises also the other heavy systems (not shown) and it recalls closely the experimental results for peripheral relativistic heavy-ion collisions~\cite{Schmidt2002,Henzlova2008}.
%
%

	We may also suggest that these results, in good agreement with previous studies based on statistical approaches, indicate that the transport description is well adapted to follow the reaching of equilibrium conditions, through a chaotic population of the available phase space,  
within the dynamical process~\cite{Raduta2006}. 
	Up to this stage, this analysis agrees with inclusive data and statistical simulations, but it is not sufficient to characterise the mechanism: both fission and multifragmentation can in fact populate the neutron-rich side of the nuclide chart due to the curvature of the $\beta$-stability valley.

\subsection{Fragmentation in few IMF}

\begin{figure}[b!]\begin{center}
	\includegraphics[angle=0, width=1\columnwidth]{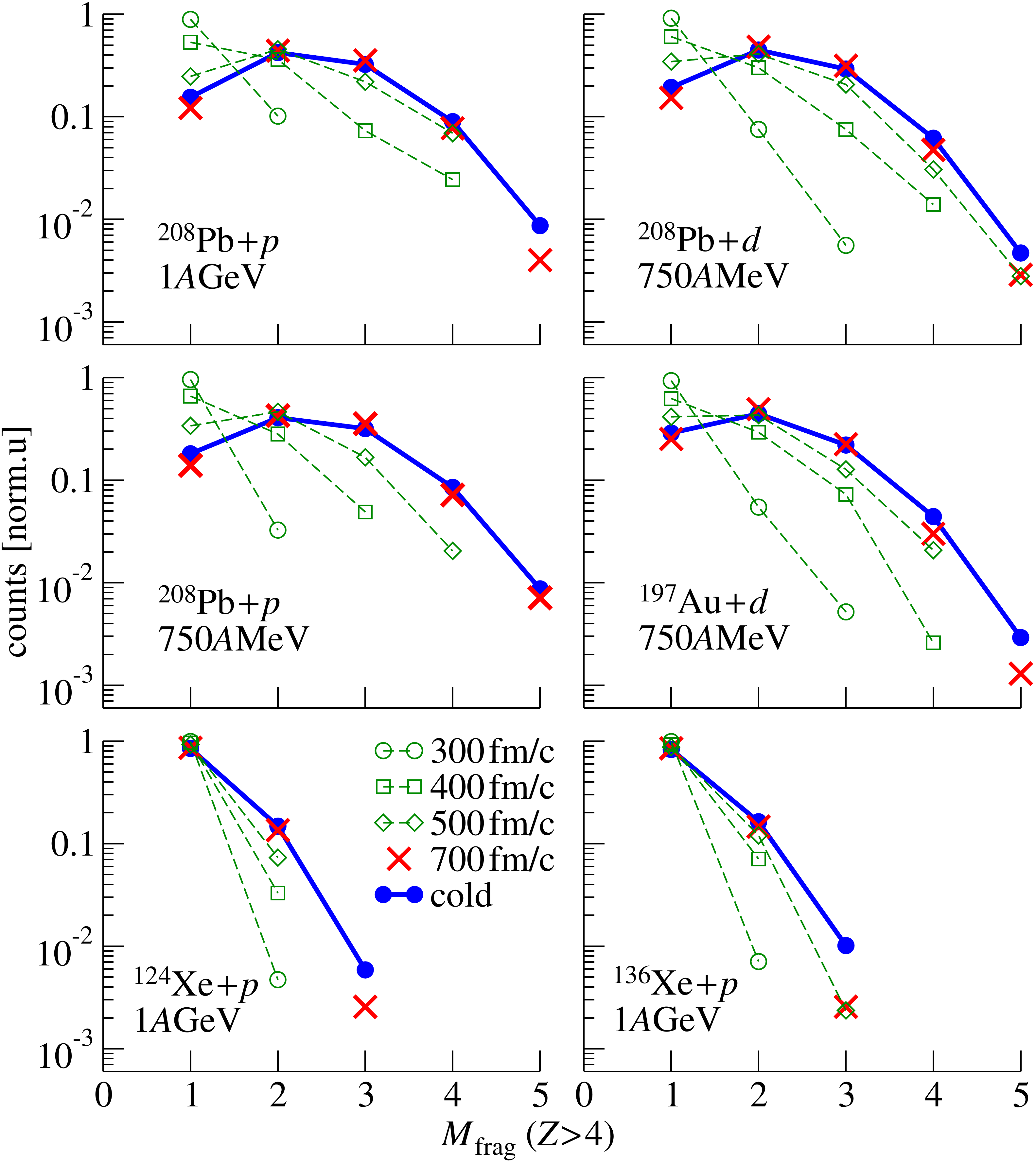}
\end{center}
\caption{
	Normalised yields as a function of the multiplicity of fragments with $Z>4$, for central impact parameters, at various time-steps of the dynamical process, and after secondary-decay.
}
\label{fig_multiplicity}
\end{figure}

	From the analysis of the multiplicity of fragments with $Z>4$ at 700 fm/c, studied for central impact parameters, we found that the lighter systems (Xe) prevalently recompact into one compound nucleus, or they undergo binary splits with about one order of magnitude smaller probability, and multiple splits are rare. 
	The heavier systems, despite also displaying some tendency to recompacting, are on the other hand dominated by binary splits, and ternary splits are also relevant.
	This analysis is presented in fig.~\ref{fig_multiplicity}.
	The evolution of the fragment-multiplicity spectrum is also shown as a function of time: we observe that, even if density inhomogeneities arise at earlier times, the system starts separating into fragments rather late, at around 300~fm/c.

%
%
%
%
%
\begin{figure}[t!]\begin{center}
	\includegraphics[angle=0, width=1\columnwidth]{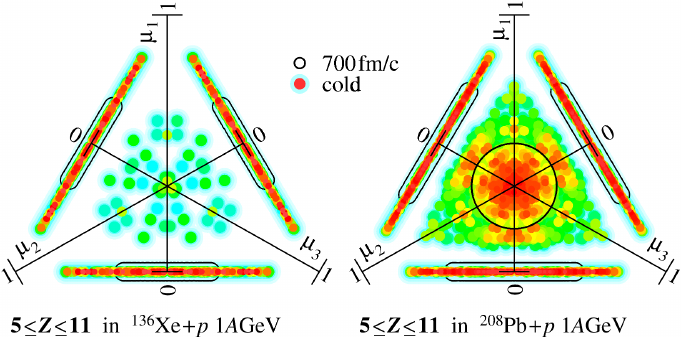}
\end{center}
\caption{
	Correlation among the three heaviest IMF's of mass number $A_1$, $A_2$ and $A_3$ produced in the same event, for all events where at least two fragments are found in the range $5\le Z\le 11$.
	Combinations of relative sizes $\mu_1$, $\mu_2$ and $\mu_3$ (where $\mu_i=A_i/(A_1+A_2+A_3)$) are studied in a Dalitz plot for the systems  $^{136}$Xe and $^{208}$Pb, for central impact parameters.
	Colour maps refer to cold systems after secondary-decay and the configurations at 700fm/c are indicated by black contour lines.
}
\label{fig_Dalitz}
\end{figure}

	An insight about the asymmetry of the splits is proposed in fig.~\ref{fig_Dalitz} by analysing the size correlation among the three heaviest IMF's, of mass number $A_1$, $A_2$ and $A_3$, produced in the same event, for events where at least two fragments are found in the range $5\le Z\le 11$.
	All combinations of the relative sizes $\mu_1$, $\mu_2$ and $\mu_3$, where $\mu_i=A_i/(A_1+A_2+A_3)$, are used as coordinates in Dalitz plots.
	The size correlations are investigated for the systems $^{136}$Xe and $^{208}$Pb, for central impact parameters, both for the hot (at 700fm/c) and for the cold systems.
	From this analysis we infer that, even when the fragment multiplicity is larger than two, the splits exhibit a large asymmetry.
	In the $^{136}$Xe hot system, represented by black contours positioned on the sides of the plot, at maximum two IMF's are found in the range $5\le Z\le 11$, and the fragment multiplicity is completed by a heavier residue.
	The action of the secondary decay may turn some few events into three-IMF patters which enter the selection and fill the centre of the plot.
	In the $^{208}$Pb system, hot and cold, symmetric splits are still rather rare with respect to events where one heavier fragment is present.
	The configuration of the splits has an obvious consequence on the kinematics.

%
%
%
%
\begin{figure}[b!]\begin{center}
	\includegraphics[angle=0, width=1\columnwidth]{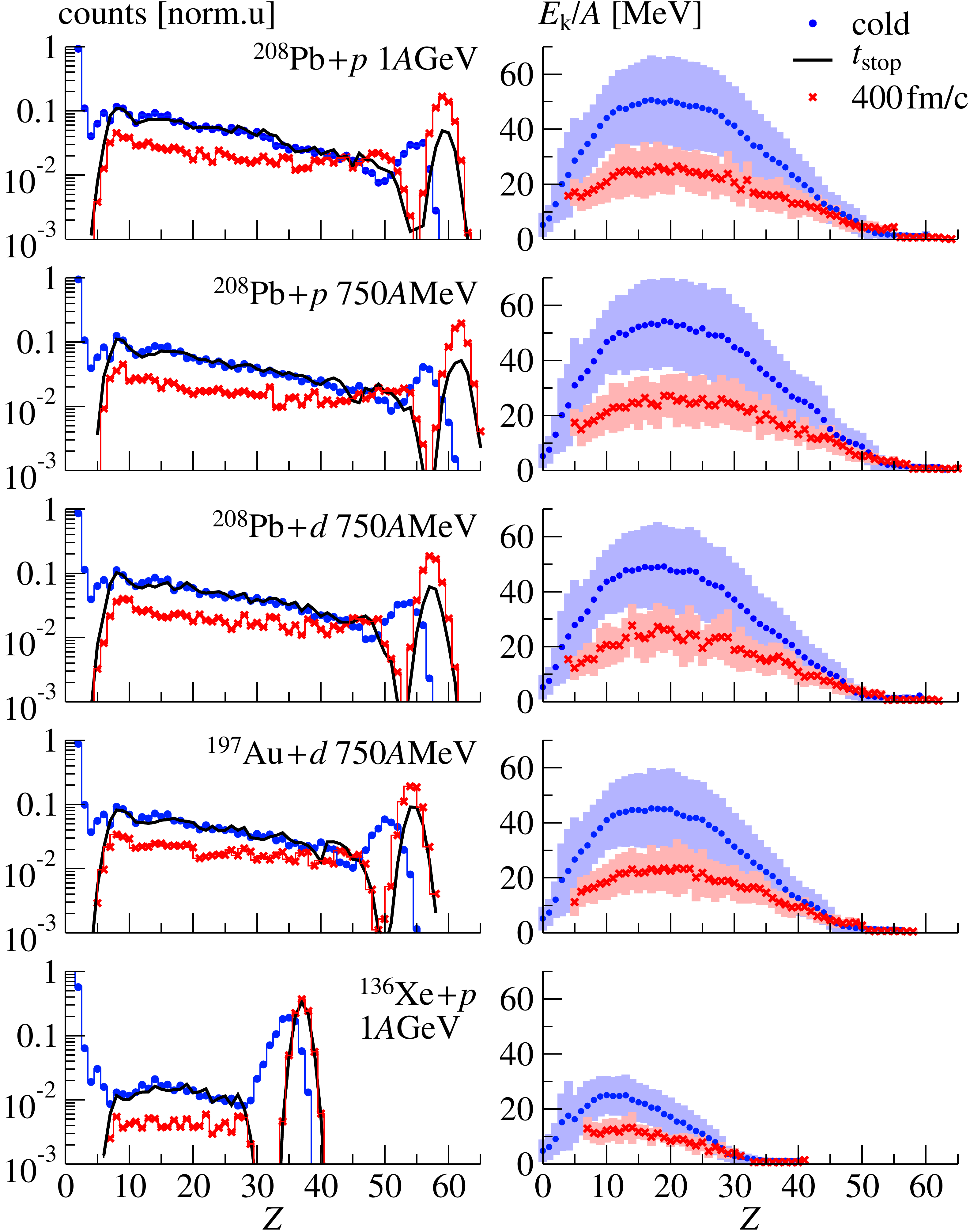}
\end{center}
\caption{
Production yields and kinetic energies for five spallation reactions, calculated as a function of the mass number, for central impact parameters, at 400fm/c, at the time $t_{\textit{stop}}$, and after secondary-decay (cold).
}
\label{fig_Adistr}
\end{figure}

\subsection{Charge distribution and kinematics: \\two emission modes for IMF's}

	The fragment-mass yields are shown in fig.~\ref{fig_Adistr}.
	The spectra at $t_{\textrm{stop}}$ and at the end of the sequential decay are similar except for the extremities, corresponding to the lightest and the heaviest masses, which have been modified by a prominent light-particle evaporation process and by asymmetric fission.
It is interesting to notice that the heavy-residue region is already filled at $t\!=\!400~fm/c$, whereas IMF's are also produced at later times. Moreover, their final yield, after de-excitation has been considered, is quite close to the yield given by the BLOB simulations at $t_{\textrm{stop}}$.    
	Therefore, within our calculation, the kinematics of the cold IMF's should mostly reflect the kinematics of the hot IMF's, when they are related to the most violent entrance channels.
	As shown in fig.~\ref{fig_Adistr}, the kinematics reveals therefore the explosive character of the process and is then modified by the Coulomb propagation.

\begin{figure}[t!]\begin{center}
	\includegraphics[angle=0, width=1\columnwidth]{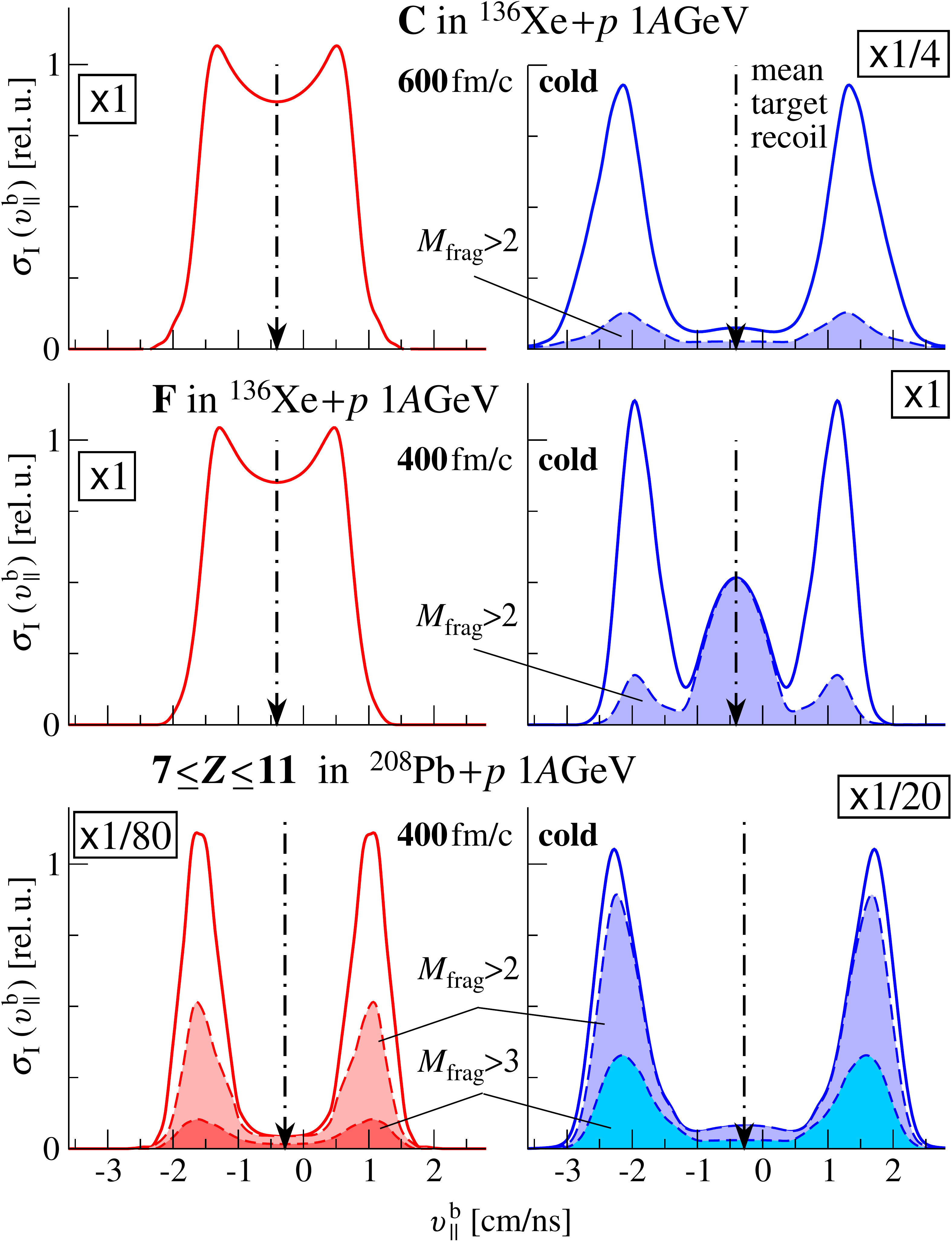}
\end{center}
\caption{
%
	Zero-angle invariant velocity distributions $\sigma_{\textrm{I}}(v_{||}^{\textrm{b}})/\sigma$  of carbon and fluorine isotopes calculated for the system $^{136}$Xe+$p$ at 1 $A$ GeV and of IMF's with $7\!\le\!Z\!\le\!11$  calculated for $^{208}$Pb+$p$ at 1 $A$ GeV.
	Left panels present distributions of hot fragments at 700 fm/c, while right panels present distributions of cold fragments after secondary decay.
	The integral of the distributions reflects the same number of events exploited for the two systems and, for better comparison, they are scaled by the factors indicated in the boxes. The spectra are shifted with respect to zero by the mean recoil velocity of the  $^{136}$Xe and $^{208}$Pb (indicated by arrows).
	The values of $M_{\textrm{frag}}$ indicate the multiplicity of fragments of $Z>4$ (including heavy residues) associated to the events.
The contribution for different fragment multiplicities is indicated.
}
\label{fig_v_BLOB}
\end{figure}

	We conclude the analysis by recalling the initial inspiring experimental finding of fig.~\ref{fig_v_distr}. 
	Due to the computational complexity, we could not collect enough statistics to reproduce the same kinematic observable of fig.~\ref{fig_v_distr} for single isotopes, but we could produce a similar observable by collecting, for instance, the velocity distributions of all isotopes of carbon and fluorine for the system $^{136}$Xe+$p$ at 1 $A$ GeV, and all IMF's with $7\!\le\!Z\!\le\!11$ (interval chosen around oxygen and neon, which are elements frequently produced in multifragmentation) for the system $^{208}$Pb+$p$ at 1 $A$ GeV: this study is illustrated in fig.~\ref{fig_v_BLOB} in the reference of the heavy nucleus before the collision; the shift with respect to zero corresponds therefore to the mean recoil of the target.
	The spectra could be symmetrised because the global reaction configuration studied with a central impact parameter is symmetric.

%
%

Carbon and fluorine in the $^{136}$Xe system change from a wide-hump distribution, for the hot IMF's, to a concave distribution for the cold IMF's. 
	When present, the contribution of events with fragment multiplicity larger than two are indicated.
For the $^{136}$Xe system it appears only in the cold system, leading to a wide convex portion of the spectrum, especially in the fluorine case: the convexity results from the variety of possible sizes and patterns involved in the splitting configurations, mainly when $M_{\textrm{frag}}>2$. 
	The resulting overall concave or two-humped wide spectra of the cold carbon and fluorine isotopes is produced by imparting different boosts to the fragments issued of binary events as a function of the partner size, producing wide humps from the folding of different Coulomb boosts, and by an additional contribution from asymmetric fission of the heavy residues, which selects a narrower Coulomb peak.
	From the analysis of fig.~\ref{fig_Dalitz} we infer that, even when the multiplicity is larger than two, the kinematics of the splits should however manifest a binary-like character due to the size asymmetry among fragments: the kinematics reflects in this case the prominent Coulomb repulsion imparted by the largest fragment. 
	This effect becomes dominant in the $^{208}$Pb system, where concave wide spectra are also observed for larger fragment multiplicities.
	The calculation was limited to a small interval of impact parameters. 
The extension to the full range of impact parameters would, firstly, add or enhance the feeding of Coulomb peaks in the cold-IMF spectrum from asymmetric fission of heavy residues.
	Secondly, it would produce a folding over a span of recoil velocities for the target. Events where IMF are produced are related to a large range of central to semi-central impact parameters, and are mostly contributing to the centre of the distribution. Thus such folding would deform the central portion of the spectrum into an asymmetric shape with more extended tails for negative values of the velocity.
	In general, we observe that the more or less pronounced filling of the centre and the appearing of wide humps in the zero-angle spectra signs the presence of mechanisms possibly related to the sudden production of a few IMF's in a same short interval of time, as suggested in experimental observations~\cite{Napolitani04,Napolitani2011}

\section{Phenomenology}

	Even though a quantitative comparison with experimental data is beyond the purpose of this work, we observe that the model describes a large range of observable, from nuclide production to kinematics, which is globally consistent with the available experimental information.
	
The dynamical evolution leads to a chaotic population of the available phase space, which makes the final result quite similar to the
predictions of statistical multifragmentation models \cite {Raduta2006} (statistical investigations along this line can be found in refs~\cite{Souza2009}).
	The interest of the dynamical approach is that it can be used to extract further information on the phenomenology of the process at any time. 
	Moreover, kinematical effects connected to the expansion dynamics can only be described within a dynamical model. 

\subsection{Frustrated multifragmentation}

%
%
\begin{figure}[t!]\begin{center}
	\includegraphics[angle=0, width=1\columnwidth]{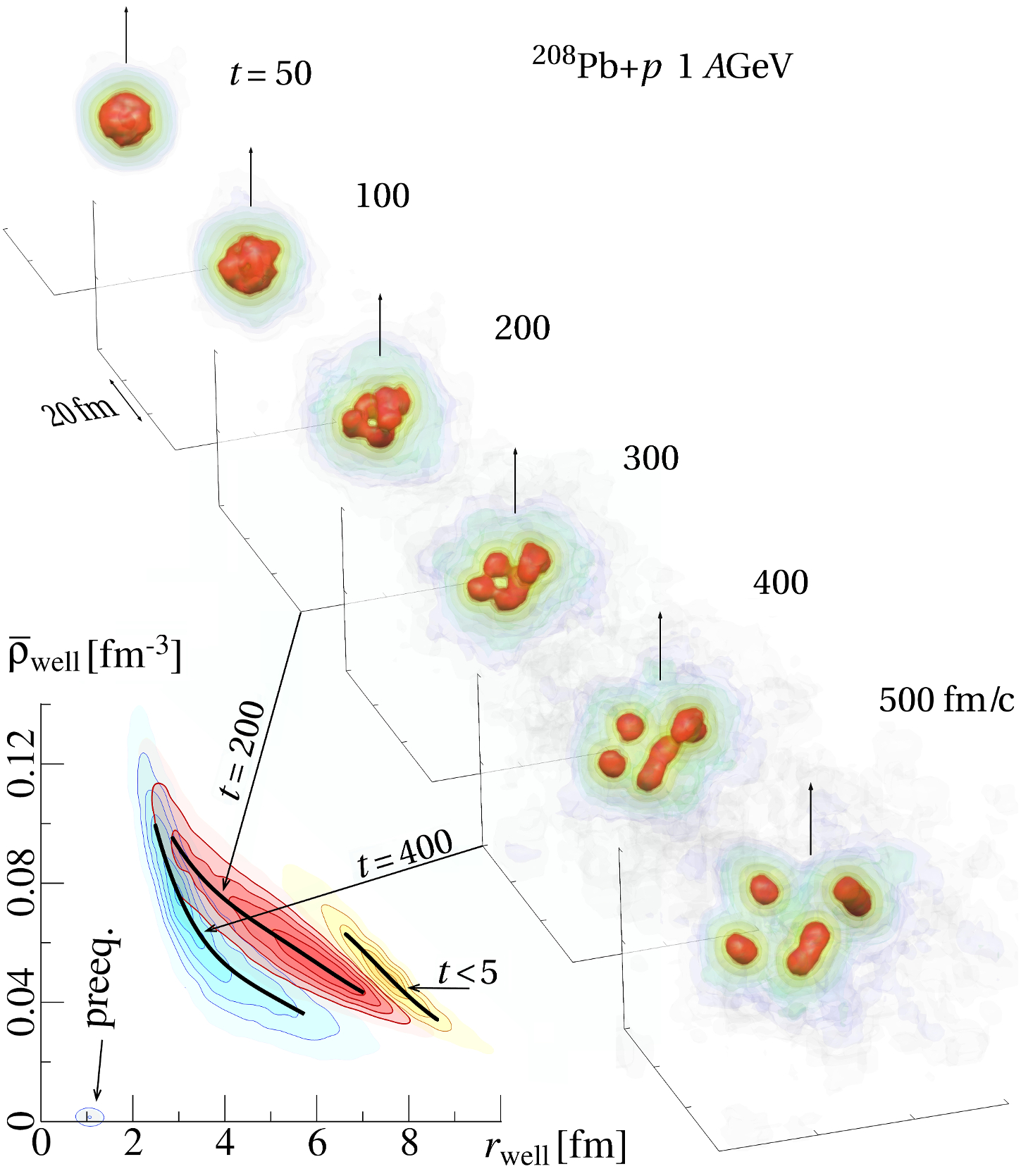}
\end{center}
\caption{
	Top. Time evolution of the density profile of the systems $p$~(1GeV)$+^{208}$Pb for one specific event selected among those giving multifragmentation.
	The system undergoes a spinodal behaviour, visible from 100 to 200~fm/c, when developing inhomogeneities of comparable size.
	Later on, the fragmentation mechanism is frustrated by the mean-field resilience, resulting into a rather asymmetric fragmentation.
	Bottom. Time evolution of the size of potential concavities associated to the evolution of the density profile at the beginning of the process ($t\!<\!5$~fm/c), during the phase of instability growth ($t\!=\!200$~fm/c), and when fragments appear ($t\!=\!400$~fm/c). See text for details. 
}

\label{fig_animation_frag}
\end{figure}

	Fig.~\ref{fig_animation_frag} gives an overview of the richness of the dynamic behaviour.
	In the first instants which follow the collision, low-density tails appear in correspondence with the emission of preequilibrium particles, proceeding from forward angles and later extending more isotropically  to all solid angles.

	After this time, the system starts expanding and the dynamical fluctuations handled by the BLOB treatment become a dominant mechanism in the process.
	With reducing bulk density, phase-space fluctuations grow in amplitude and potential ripples develop, becoming the nesting sites of fragments: inhomogeneities in the bulk density profile stand out at around 100 fm/c, but it takes them long time to eventually separate into fragments.
	The process exhibits a typical characteristic of the spinodal instability, i.e. the arising of blobs of similar size in the bulk.
	The inset of fig.~\ref{fig_animation_frag} analyses for this same system the density averaged over those blobs as a function of their size at different times.
	The blobs are identified as any potential concavity found in the system and the size 
is their average radius (their shape is nearly spherical).
	At early times potential concavities coincide with the whole expanding system or with some large portions of it when particle flow develops. 
	At late times potential concavities have a large probability to coincide with the inhomogeneities arising in the density landscape: because their size reflects the leading instability mode~\cite{Napolitani_IWM2014}, they are all comparable in size, corresponding with larger probability to neon or oxygen nuclei~\cite{Chomaz2004,Borderie2008}.
	At intermediate times, sizes range from the whole system to the size of the spinodal undulations in the density landscape.
	This coexistence recalls phase-transition signals and corresponding results at Fermi energies, where suitable 
observables (such as the asymmetry between the charges of the two heaviest fragments produced in one collision event~\cite{Pichon2006} or the size of the largest fragment produced in one event~\cite{Bonnet2009}) have been proposed. 

	However, only in presence of a sufficiently large radial expansion these blobs can separate into fragments of comparable size and preserve the spinodal signal also in the exit channel.
	We can observe that this is definitely not the case: not all blobs succeed in separating in single fragments but they bond together in groups.
	The event of fig.~\ref{fig_animation_frag} finally results in the fragmentation of the system into four asymmetric parts.
	Such a scenario seems to be general for this kind of spallation mechanisms and it is reflected in the low multiplicity of hot fragments analysed in fig.~\ref{fig_multiplicity} and in the mass distribution of hot remnants of fig.~\ref{fig_Adistr}. 
	In fact this latter, even when displaying a peak around the elements selected by the spinodal instability like oxygen and neon, presents a rather flat distribution which implies a large recombination of the spinodal inhomogeneities into larger fragments.
As it was already argued in Ref.\cite{Colonna1997}, we can conclude on this phenomenology that the multifragmentation mechanism in spallation in the 1~$A$GeV energy regime is frustrated by the action of the mean field which tends to recompact the system as far as not enough energy is spent in the radial expansion.
%

\subsection{Exit-channel chaos and binary events}

%
%
\begin{figure}[b!]\begin{center}
	\includegraphics[angle=0, width=.8\columnwidth]{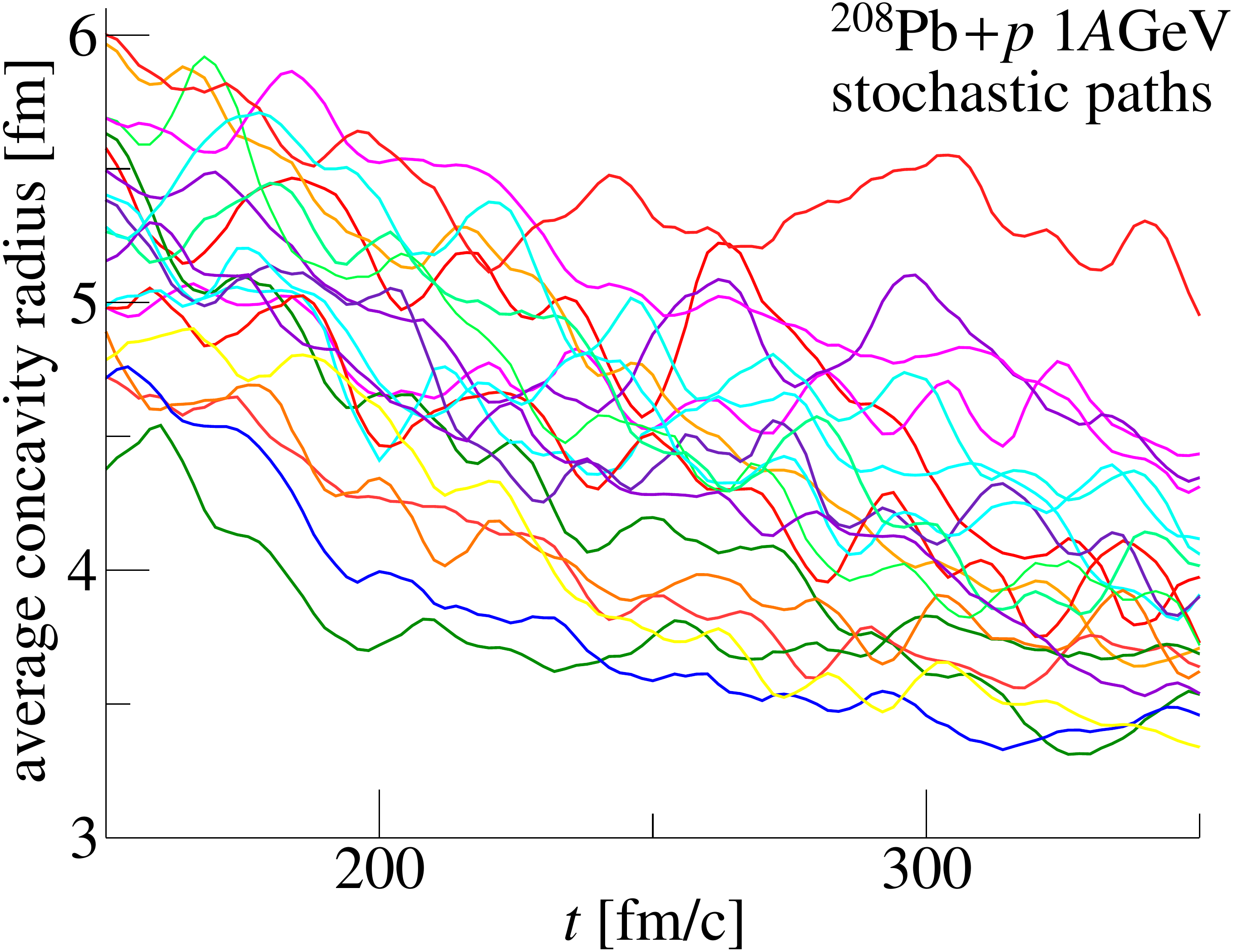}\\
\end{center}
\caption{
	A random selection of some reaction paths followed by the system $^{208}$Pb+$p$ at 1 $A$ GeV on the time-dependent potential landscape, represented by the average size of potential concavities as a function of time.
}
\label{fig_trajectories}
\end{figure}

	Phase-space fluctuations favour fragmentation.  Moreover,
	they also act, over many events, in expanding the bundle of dynamical reaction paths into a large chaotic pattern of bifurcations~\cite{Randrup1990}:
this leads to a variety of exit channels.
	Fig.~\ref{fig_trajectories} illustrates this effect by plotting the average radius of inhomogeneities found in the density landscape as a function of time for several stochastic evolutions of the systems $^{208}$Pb+$p$ at 1 $A$ GeV.
	Each trajectory could lead to a different exit channel, characterised by a different kinematics and larger or smaller fragment multiplicities and charge asymmetries.

\begin{figure}[b!]\begin{center}
	\includegraphics[angle=0, width=.99\columnwidth]{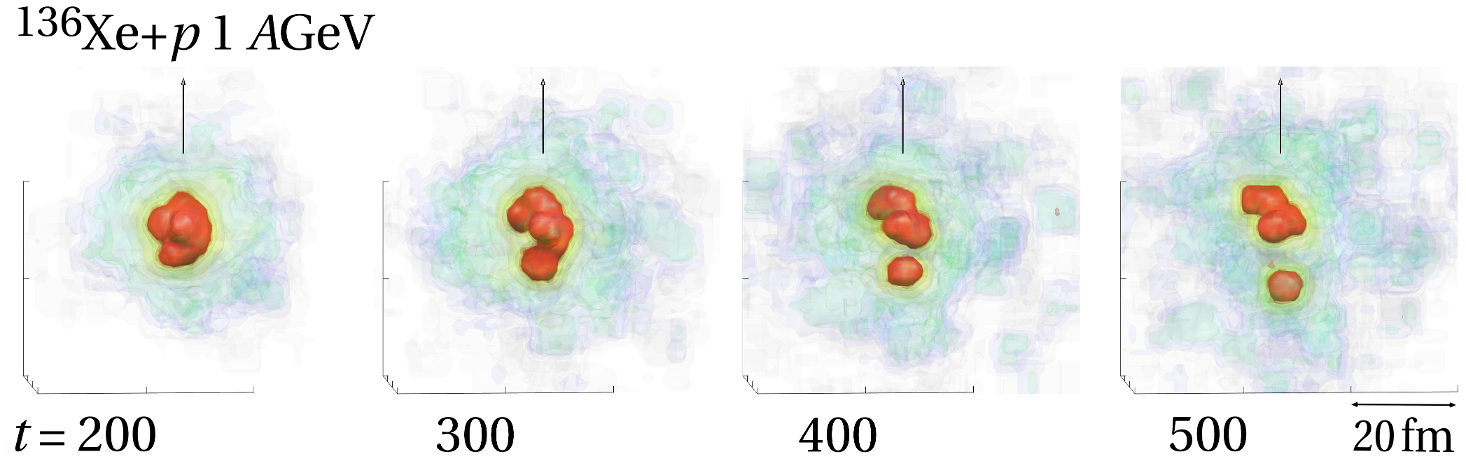}
\end{center}\caption
{
	Time evolution of the density profile for the systems $^{136}$Xe+$p$ at 1 $A$ GeV for one events resulting into an asymmetric binary split.
}
\label{fig_animation_fiss}
\end{figure}

	Particularly interesting are trajectories leading to only two fragments in the exit channel, as it can occur rather often in systems like $^{136}$Xe+$p$ at 1 $A$ GeV, as illustrated in fig.~\ref{fig_animation_fiss}.
	Also in this case, a spinodal process may activate and immediately enter in competition with  the action of the mean field which tends to reverse the fragmentation pattern into a compact shape.
	Most of the times, this frustrating process results in one single compound nucleus.
	Very seldom the mean field succeeds only partially in coalescing the inhomogeneities of the density profile, and one or a group of those separates into a fragment and leaves the system.
	Depending on the stochastic configuration of the fragmenting system, the partial coalescence could recompact the inhomogeneities in many combinations resulting in different asymmetries.

%
%
%
%

	To characterise the final exit channel, we have combined the dynamical process and the additional secondary decay, which we simulated through a fission-evaporation afterburner.
	It turns out that two mechanisms may coexist in the spallation systems we examined.
	One is a multifragmentation process frustrated by the coalescence effect of the mean field, which constrains the fragment multiplicity to small values, rarely equal to three or four, more often equal to two: this work focused mostly on this process.
	The other mechanism, related to the production of heavy residues, is also a binary split, but it coincides with asymmetric fission from a compound nucleus in the secondary decay process~\cite{Sanders99}.
	For different causes, both processes contribute to the same IMF production when regarding the charge distribution and even when regarding the isotopic content.
	However, these two contributions are fundamentally different because of two reasons.
	First: They are separate in time; multifragmentation develops within the short time of the dynamical collision process, while the second occurs during the longer time of the fission decay process.
	Second: from a dynamical point of view  binary splits emerging from fission or from multifragmentation are different: fission of a compound nucleus is a trajectory in a deformation landscape which passes through the development of a neck; on the other hand, multifragmentation leads initially to a mottling topology in density space, possibly driven by spinodal instabilities, and then the expansion dynamics attempts to keep this topology, in competition with the antagonist tendency of reverting into a compact shape.
This would lead to binary channels, similar to fission, but obtained through re-aggregation processes. The latter could generally have
different kinematic features with respect to standard fission processes.  
	Experimentally, this difference would be reflected in the velocity spectra of the IMF's if the resolution is sufficiently high.

	The length of the transport calculation makes it prohibitive to track all contributions to the yields coming from the full distribution of excitation energies of the spallation system, and we had to restrict to the most violent collisions only.
	More quantitative simulations are left for further works.

\section{Concluding remarks}
	The Boltzmann-Langevin one body (BLOB) approach has been applied to the dynamics of hot nuclear systems produced in spallation reactions.
	A simplified cascade procedure was introduced to simulate the collision of relativistic protons and light projectiles on heavy target nuclei.
	Successively, the dynamics is followed in time within the BLOB treatment, to investigate the trajectory followed by the system as a function of the initial energy deposition. 

	We find that unstable isoscalar modes can actually arise in such systems, like heavy nuclei bombarded by protons and deuterons in the 1 A GeV regime and, with a low but not negligible probability, become responsible for the fragmentation of the system.
	According to our theoretical approach, we find that unstable modes in spallation should exhibit quite a similar phenomenology as spinodal instability in dissipative central heavy-ion collisions when the incident energy corresponds to the threshold between fusion and multifragmentation \cite{Napolitani2013}.

	In the case of dissipative ion-ion collisions, the excitation of the system is mostly determined by mechanical perturbations, while in the spallation process the excitation originates from an almost isotropic propagation of the energy deposited by the light projectile.
	Though these situations are different in some aspects, they both drive phase-space fluctuations of large amplitude and they may both activate the spinodal behaviour and the amplification of mechanically unstable modes.
	In both cases, the mean field may then have the effect of reverting the whole system, or part of it, into a compact shape, smearing, modifying, or completely erasing the fragment configuration.
	In particular, if the kinetic energy feeding the expansion dynamics is not sufficient to disintegrate the system but it is still larger than the system can hold, the fragment multiplicity reduces and the fragment configuration becomes asymmetric.
	As an extreme situation, the process may look like asymmetric fission, but the chronology, as well as the violence of the process will be incompatible with the conventional fission picture: this process would correspond to a binary channel obtained by re-aggregation, keeping some dynamical aspects.  

	Further binary contribution to IMF deriving from fission have been accounted for by adding to the dynamical simulation a fission-evaporation treatment for the decay of all the possible warm configurations explored by the collision.
	The overall simulation, constructed around a Boltzmann-Langevin description of the most violent collision events, leads to a correct qualitative picture for the production and kinematics of IMF nuclides. 
	In particular, the physical description that we suggest for IMF production in spallation may solve apparent discrepancies between some experimental interpretations of inclusive and exclusive data, the former revealing multifragmentation-like kinetic energies, the latter revealing small fragment multiplicities compatible with compound-nucleus decays.
	These experimental results would actually be perfectly coherent.

	As a conclusion, we found within a dynamical transport approach incorporating a Langevin term that the incident energy of a light projectile in the 1 GeV range is sufficient to turn the heavy target into an unstable system, where mechanically unstable modes develop leading to a variety of aymmetric fragment configurations, including binary channels. 
	This phenomenology reflects the entrance of the system in the spinodal zone of the nuclear matter phase diagram and appears as a frustrated spinodal behaviour.

\section{Aknowledgements}
	Reasearch conducted in the scope of the International Associated Laboratory (LIA) COLL-AGAIN.

%
%
%
%

%
%
%
%
\end{document}